\documentclass[fleqn,12pt]{wlscirep}

\usepackage[utf8]{inputenc}
\usepackage[T1]{fontenc}
\usepackage{lineno}
\linenumbers
\usepackage{color,soul}
\usepackage{upgreek}
\usepackage{mathastext}
\usepackage{setspace}
\usepackage{eqnarray,amsmath}

\nolinenumbers
\title{\centering  Physics-inspired Neuroacoustic Computing Based on Tunable Nonlinear Multiple-scattering }

\author[1]{Ali Momeni}
\author[2]{Xinxin Guo}
\author[2]{Hervé Lissek}
\author[1,*]{Romain Fleury}
\affil[1]{Laboratory of Wave Engineering, Ecole Polytechnique Fédérale de Lausanne (EPFL), CH-1015 Lausanne, Switzerland.}
\affil[2]{Signal Processing Laboratory LTS2, Ecole Polytechnique Fédérale de Lausanne (EPFL), CH-1015 Lausanne, Switzerland}
\affil[*]{E-mail: romain.fleury@epfl.ch}
\doublespace 

\begin{abstract}
 Waves, such as light and sound, inherently bounce and mix due to multiple scattering induced by the complex material objects that surround us. This scattering process severely scrambles the information carried by waves, challenging conventional communication systems, sensing paradigms, and wave-based computing schemes. Here, we show that instead of being a hindrance, multiple scattering can be beneficial to enable and enhance analog nonlinear information mapping, allowing for the direct physical implementation of computational paradigms such as reservoir computing and extreme learning machines. We propose a physics-inspired version of such computational architectures for speech and vowel recognition that operate directly in the native domain of the input signal, namely on real-sounds, without any digital pre-processing or encoding conversion and backpropagation training computation. We first implement it in a proof-of-concept prototype, a nonlinear chaotic acoustic cavity containing multiple tunable and power-efficient nonlinear meta-scatterers. We prove the efficiency of the acoustic-based computing system for vowel recognition tasks with high testing classification accuracy ($91.4\%$). Finally, we demonstrate the high performance of vowel recognition in  {the natural environment of a} reverberation room. Our results open the way for efficient acoustic learning machines that operate directly on the input sound, and leverage physics to enable Natural Language Processing (NLP).

\end{abstract}
\begin{document}
\flushbottom
\maketitle

\thispagestyle{empty}
\section{Introduction}

Wave  propagation in a complex and inhomogeneous linear medium can be seen as a linear mixing process between a set of input and output modes. The number of spatial degrees of freedom that can carry information is related to the ratio of the size of the system to the wavelength, which can be extremely large in optics \cite{gigan2022imaging,cao2022shaping,zangeneh2021analogue,wright2022deep}. Performing wave-based analog computing and signal processing tasks such as matrix multiplications, convolution, image differentiation, and random projections, has attracted a surge of interest due to the possibility of operating with very low-power, in massively parallel and large-scale architectures, without the need for memory as in conventional digital computers \cite{feldmann2021parallel,silva2014performing, wang2022optical,momeni2021reciprocal,sol2022meta,babaee2021parallel,del2018leveraging,matthes2019optical,momeni2019generalized,momeni2022switchable,anderson2023optical,wetzstein2020inference,wu2020neuromorphic,lin2018all}. In fact, all current computer architectures, from conventional computers to specialized architectures for machine learning, such as graphical processing units (GPUs) or Google's tensor processing units (TPUs), suffer from the so-called von Neumann's bottleneck: the separation of memory and processing unit is associated with a considerable penalty in terms of speed and energy consumption \cite{sebastian2020memory}. On the other hand, analog computing using waves belongs to a distinct class of computing paradigms known as neuromorphic computing, which is unaffected by these shortcomings \cite{markovic2020physics,grollier2020neuromorphic}.

\begin{figure}[h!]
	\centering
	\includegraphics[width=17.1cm]{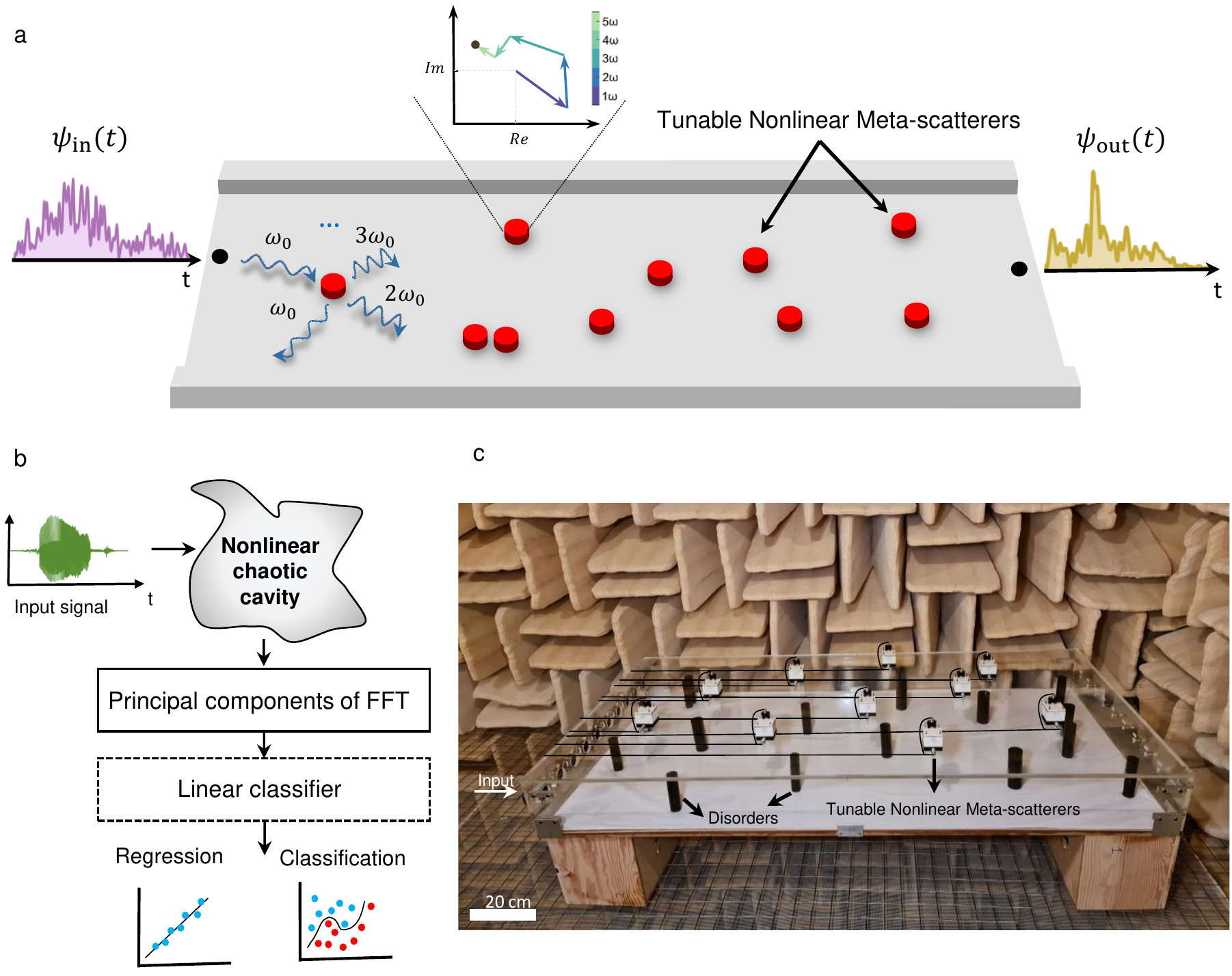}
	\caption{\textbf{ Physics-inspired reservoir computing.} \textbf{a}, Schematic of a physical nonlinear chaotic computing system and \textbf{b},  corresponding block diagram. Raw waveforms $\psi_{in}(\textit{t})$ are injected at the left input, and propagate through a nonlinear chaotic cavity. As the waves propagate in the cavity, they encounter tunable nonlinear meta-scatterers, which scatter the wave into higher harmonics. The output $\psi_{out}(\textit{t})$ is fed into an adaptable dense layer (linear classifier) with rescaled coefficients, and used for regression and classification.  \textbf{c}, Experimental implementation of the proposed computing system with sound waves and an acoustic cavity including fifteen  sub-wavelength rods and ten nonlinear meta-scatterers, in the form of nonlinear active membranes, randomly wall-mounted about the cavity.   
}\label{Fig_1}
\end{figure}

The analogy between wave propagation in inhomogeneous and/or nonlinear media and a single-layer neural network can be used to directly process the information carried by waves. Such analogies between wave-physics and computer science have nurtured a great variety of concepts in both fields, for instance, the analogy between Nondeterministic Polynomial time, NP-hard, problems and photonic Ising machines \cite{pierangeli2019large,huang2021antiferromagnetic}.
Wave propagation in complex and/or nonlinear media opens up the ability to perform random or nonlinear kernels for different learning tasks with a speed independent of the number of neurons,  by which a simple linear Support Vector Machine (SVM) \cite{suykens1999least} or a linear regression  \cite{seber2003linear} can classify the transformed data, obtained through a complex medium (such as in wave-based reservoir computing (RC) and extreme learning machines (ELM)) \cite{rafayelyan2020large,pierangeli2021photonic,teugin2021scalable,pathak2018model,momeni2022electromagnetic,yildirim2022nonlinear,oguz2022programming,vandoorne2014experimental,zhou2021large}. However, the required nonlinearity to realize such wave-based RC and ELM systems is generally implemented on the digital front of the wave-based system, or introduced by intrinsic material nonlinearity (such as in Kerr dielectrics), which impairs energy efficiency, flexibility, and speed \cite{yildirim2022nonlinear,teugin2021scalable,hughes2019wave,skinner1994optical}.

On the other hand, the possible advantages of nonlinear strongly scattering media in wave-based neuromorphic computing have been left largely uncharted. Multiple scattering is typically associated with strong spatial-dispersion, namely an entanglement between the response of distant points in a medium, which could potentially boost the interconnectivity of signal flow, intensify the nonlinearity, and enable going way beyond the single-layer model. This provides an opportunity to process information in a real-world environment by only adding some nonlinear meta-scatterers in an already disordered medium, which would be particularly relevant in the realms of radio-frequency or acoustic waves, for which cavities and nonlinear meta-scatterers are readily available. 

Here, we demonstrate a wave-based neuromorphic reservoir computer that performs fast and power-efficient nonlinear data transformations directly on the wave signal to be sensed, eliminating conversion or pre-processing and heavy training computation of backpropagation \cite{rumelhart1986learning,wright2022deep}. We demonstrate the relevance of such a scheme for a challenging vowel recognition problem, solved by letting sound propagate through a multiple-scattering cavity doped with nonlinear meta-scatterers, and by processing the output of a few microphones with a simple linear classifier.
We experimentally investigate how nonlinear meta-scatterers randomly located in a homogeneous medium can enhance the computing performance of a first prototype, a nonlinear acoustic cavity. We prove the efficiency of the acoustic-based computing system for challenging vowel recognition tasks, obtaining a high classification accuracy for test sounds ($91.4\%$). We then demonstrate the flexibility of our scheme by showing the high performance of vowel recognition in a realistic scenario, in a real strongly-scattering reverberation room. Such wave-based computing systems are not only fast, easy to train, power-efficient, and versatile, but also feature a unique accuracy performance that is comparable to that obtained with the most elaborated digital schemes.

\section{Results}

\subsection{Physics-inspired neuromorphic reservoir computing}
We consider a particular type of recurrent neural network scheme, known as reservoir computing. In reservoir computing, recurrently connected neurons with fixed random weights form a reservoir and are all connected to a few output neurons via a trainable layer \cite{zhong2021dynamic}.
The connected neurons with fixed random weights can significantly simplify the training process, which is a key benefit of RC. Also, the recurrent nature of RC, makes it suitable for performing time-series prediction and classification\cite{sun2021sensor,marinella2019efficient,liu2022optoelectronic,du2017reservoir,milano2022materia,moon2019temporal}. 

We now briefly introduce the concept of conventional RC. An input vector $\mathbf{\psi}^{in}(\textit{t})$ is injected into a high-dimensional dynamical system called the “reservoir”. The vector $\mathbf{h}_\textit{t}$ represents the reservoir state and its initial state is defined randomly. The internal connections of the reservoir neurons and the connections between the input and the reservoir neurons are respectively defined by $\mathbf{W}_{in}$ and $\mathbf{W}_{r}$.
Both matrices are initialized randomly and fixed during the whole RC training process. The evolution of the neurons of a reservoir computer can be expressed as a recursive nonlinear mapping of the form \cite{tanaka2019recent}:
\begin{align}\label{e1}
    \mathbf{h}_{\textit{t}+1}=F_{NL}\bigg(\mathbf{W}_{in}\mathbf{\psi}^{in}_\textit{t}+\mathbf{W}_{r}\mathbf{h}_\textit{t}\bigg)
\end{align}
where $F_{NL}$ is a nonlinear function. In the training procedure, the input $\mathbf{\psi}_{in}$ is fed to the reservoir, and the corresponding reservoir states are recursively calculated. The final step of the training process is to perform a simple linear support vector machine or linear regression in order to minimize the loss function that adjusts the $\mathbf{W}_{out}$ weights. The output can be computed with $\mathbf{O}_\textit{t}=\mathbf{W}_{out}\mathbf{h}_\textit{t}$.  From Eq. (1), the reservoir is defined as a dynamical system with unique memory and nonlinearity properties. In such systems, the reservoir state is updated at each time step depending on its previous value, allowing the system to retain the memory of past information and learn long-range dependencies in data. 

{T}his scheme can be implemented using the nonlinear multiple-scattering neuro-acoustic system of 
\textcolor{blue}{Fig. \ref{Fig_1}a}, whose block diagram is given in \textcolor{blue}{Fig. \ref{Fig_1}b}. It consists of multiple nonlinear meta-scatterers randomly located in a homogeneous medium, for example, a cavity.  
We let the input sound $\mathbf{\psi}^{in}(\textit{t})$, injected on the left, propagate through the cavity. Linear and nonlinear meta-scatterers provide a multitude of multiple-scattering events. In our experiment, we designed {power-efficient}  active nonlinear meta-scatterers for acoustic waves, whose properties will be detailed later. 

The dynamics of the acoustic pressure $\textit{p}(\textit{x}, \textit{y}, \textit{z})$ are governed by the acoustic wave equation $\frac{\partial^{2}\textit{p}}{\partial \textit{t}^{2}}-\textit{c}^{2} \cdot \nabla^{2} \textit{p}=\mathbf{\psi}^{in}_\textit{t}$,
\begin{figure*}[thbp]
	\includegraphics[width=1\textwidth]{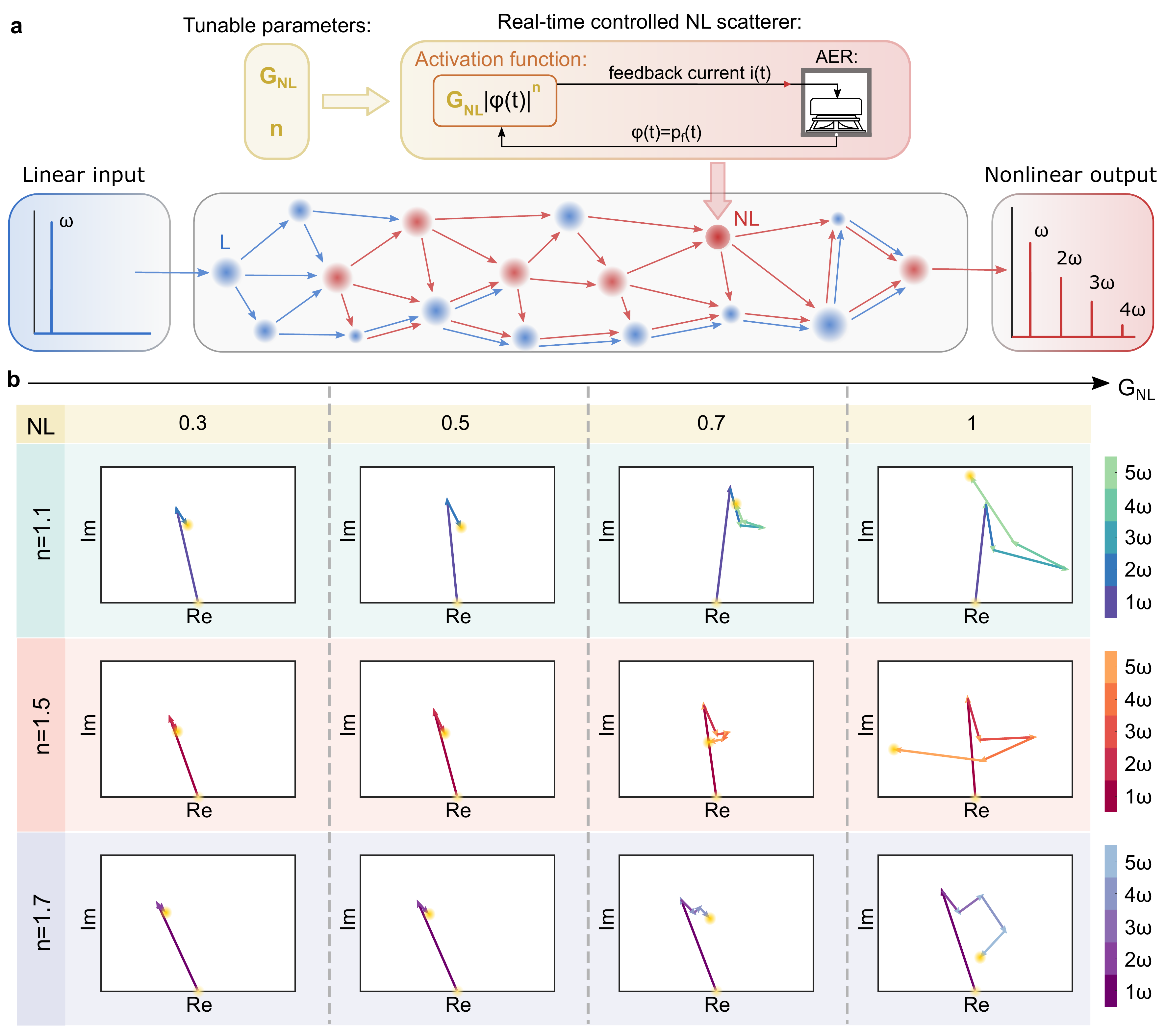}
	\centering
	\caption{\label{fig: NL} \textbf{Fully tunable nonlinear multiple scattering.} \textbf{a} Nonlinear scatterers (in red) are made of Active Electrodynamic Resonators (AER), in the form of  loudspeaker membranes controlled in real-time by a feedback current $\textit{i}(\textit{t})$, with nonlinearity specified based on the front pressure $p_f$. \textbf{b} Demonstration of the tunability of the nonlinear scattering process: stroboscopic picture (evolution of the complex phasors) of the first five pressure harmonics measured in front of a nonlinear scatterer, when varying the type and the level (defined by the gains $G_{NL}$) of nonlinearity (defined respectively by the power $\textit{n}$ and the gain $G_{NL}$). The whole system is excited by a sinusoidal wave at 500Hz. In each nonlinear case, the real (abscissa) and the imaginary (coordinate) parts of all harmonics are scaled by the absolute magnitude of the corresponding fundamental component $\omega$. The light yellow and intense yellow points show the initial location (0,0) and the final location, respectively.}\label{Fig_2}
\end{figure*} 
where $\nabla^{2}= \frac{\partial^{2}}{\partial \textit{x}^{2}}+\frac{\partial^{2}}{\partial \textit{y}^{2}}+\frac{\partial^{2}}{\partial \textit{z}^{2}}$ is the Laplacian operator, $\textit{c} = \textit{c}(\textit{x}, \textit{y}, \textit{z})$ is the spatial distribution of the wave speed, and $\psi^{in}_\textit{t} = \psi^{in}_\textit{t}(\textit{x}, \textit{y}, \textit{z}, \textit{t})$ is a source term. 
A finite difference temporal discretization of this equation results in a recurrent equation of the form
\begin{equation}\label{e2}
\mathbf{h}_\textit{t}=\mathbf{Q}^{(c)}(\mathbf{h}_{\textit{t}-1}) \cdot \mathbf{h}_{\textit{t}-1}+\mathbf{Q}^{(i)} \cdot \mathbf{x}_\textit{t}
  \end{equation}

where $\mathbf{h}_\textit{t}=\begin{pmatrix}
 \textit{p}_{\textit{t}+1} \\
\textit{p}_\textit{t} 
\end{pmatrix}$ and the sparse matrix, $\mathbf{Q}^{(c)}$, describes the update of the wave fields $\textit{p}_\textit{t}$ and $\textit{p}_{\textit{t}-1}$ without a source. The {formulation of matrices} $\mathbf{Q}^{(c)}$ and $\mathbf{Q}^{(i)}$, as well as the derivation of  Eq. \eqref{e2}, are given in Supplementary section S1.
The dependence of $\mathbf{Q}^{(c)}$ on $\textit{p}_{\textit{t}-1}$ can generally be nonlinear, leading to a nonlinear response of the wave system. Comparing Eq. \eqref{e1} {with Eq.} \eqref{e2} shows a clear similarity between conventional RC and wave systems in terms of dynamics. This recurrence relation, {expressing the} dependency of wave fields at time step $\textit{t}$ to  {the state} at time step $\textit{t}-1$, shows the intrinsic short-term memory of wave-based systems. Besides the internal memory of wave propagation through a {medium}, long-term memory is also explicitly built into the complex structure and enhanced by multiple scattering. To this end, we leverage and engineer resonant scattering elements that realize enhanced long-term memory by intrinsic wave feedback, creating complex ‘mazes’ of possible information propagation paths. The memory effect can be experimentally visualized by the impulse response when the wave-based system is excited by a Dirac delta function $\psi^{in}(\textit{t}) = \delta(\textit{t}-\textit{t}_\textit{0})$. The fact that the impact of input at time step $\textit{t}_\textit{0}$ extends into a specific time horizon serves as an illustration of the memory effect. We illustrate this result for both our nonlinear chaotic cavity and the reverberation chamber in Supplementary section S3.

\subsection{Fully tunable nonlinear multiple scattering processes}
We built a prototype of a physical wave-based system for neuromorphic computing, shown in \textcolor{blue}{Fig. \ref{Fig_1}c}. The employed acoustic cavity consists of multiple nonlinear meta-scatterers randomly placed on the cavity top wall. The temporal input signals are injected by ten loudspeakers on the right side of the cavity, emitting waveforms that propagate through a disordered 
cavity containing ten nonlinear meta-scatterers. The waveforms are received by ten microphones on the location of nonlinear meta-scatterers.

To implement the desired nonlinear meta-scatterers, an active control strategy is applied to sub-wavelength electrodynamic loudspeakers, inspired by a known method initially used in sound absorption applications \cite{etienne_IEEE, GUO_JSV_2022}. It can tailor the acoustic properties of the loudspeaker by feeding back an electrical current $\textit{i}(\textit{t})$ in real time. 

It can allow significant nonlinear effects at arbitrarily low input power \cite{GUO_PRApplied_2020}. This implies that a wave with central frequency $\omega_0$ will generate, as it propagates, spectral components at frequency multiples of $\omega_\textit{0}$.
In a view to achieving an optimal coherent energy transfer towards higher harmonics, nonlinear control laws are designed herein in the form of $\textit{G}_{\mathrm{NL}}\textit{p}_f^n$, as depicted in \textcolor{blue}{Fig. \ref{Fig_2}a}, where $\textit{p}_f(\textit{t})$ is the sensed acoustic pressure in front of the controlled loudspeaker. $\textit{G}_{\mathrm{NL}}$ and $\textit{n}$ are the two control parameters to adjust the level (with $\textit{G}_{\mathrm{NL}}$) and the form (with index $\textit{n}$) of nonlinearity, respectively. 

When nonlinearity is introduced to meta-scatterers through active means, the induced nonlinear multiple-scattering process becomes fully tunable. This can be demonstrated by measuring the temporal signal of the acoustic pressure at a given location, taking a single-sided Fourier transform, and plotting the Fourier components of each harmonic as arrows in the complex plane. Their summation allows us to visualize the value of the acoustic pressure at discrete times, giving a stroboscopic picture of the multi-frequency sound at each period of the fundamental harmonic, as represented in \textcolor{blue}{Fig. \ref{Fig_2}}. Three types of nonlinearity are considered in \textcolor{blue}{Fig. \ref{Fig_2}b}, characterized by different powers namely $\textit{n}=1.1, 1.5, 1.7$, respectively. For each type of nonlinearity, the nonlinear gain $\textit{G}_{NL}$ is used to vary the level of nonlinearity. When it is increased to the maximum value allowed by stability, the higher harmonics can reach amplitudes comparable to one of the fundamental components, which is particularly noticeable in the case of $\textit{G}_{NL}=1$. This highlights the impact of multiple scattering in a chaotic cavity, resulting in a noticeable amplification of nonlinearity, i.e., an increase in the intensity of higher harmonics, which is unattainable in a free-space scenario.  Moreover, small changes in the value of the exponent $\textit{n}$, from $\textit{n}=1.1$ to $\textit{n}=1.7$ lead to very different complex-domain profiles of all the harmonics (as can be seen in \textcolor{blue}{Fig. \ref{Fig_2}b}). This enhanced sensitivity to nonlinear parameters, due to nonlinear multiple scattering, will allow us to optimize the type and level of nonlinearity appropriate for a given computing task.

  \begin{figure}[h!]
	\centering
	\includegraphics[width=\textwidth]{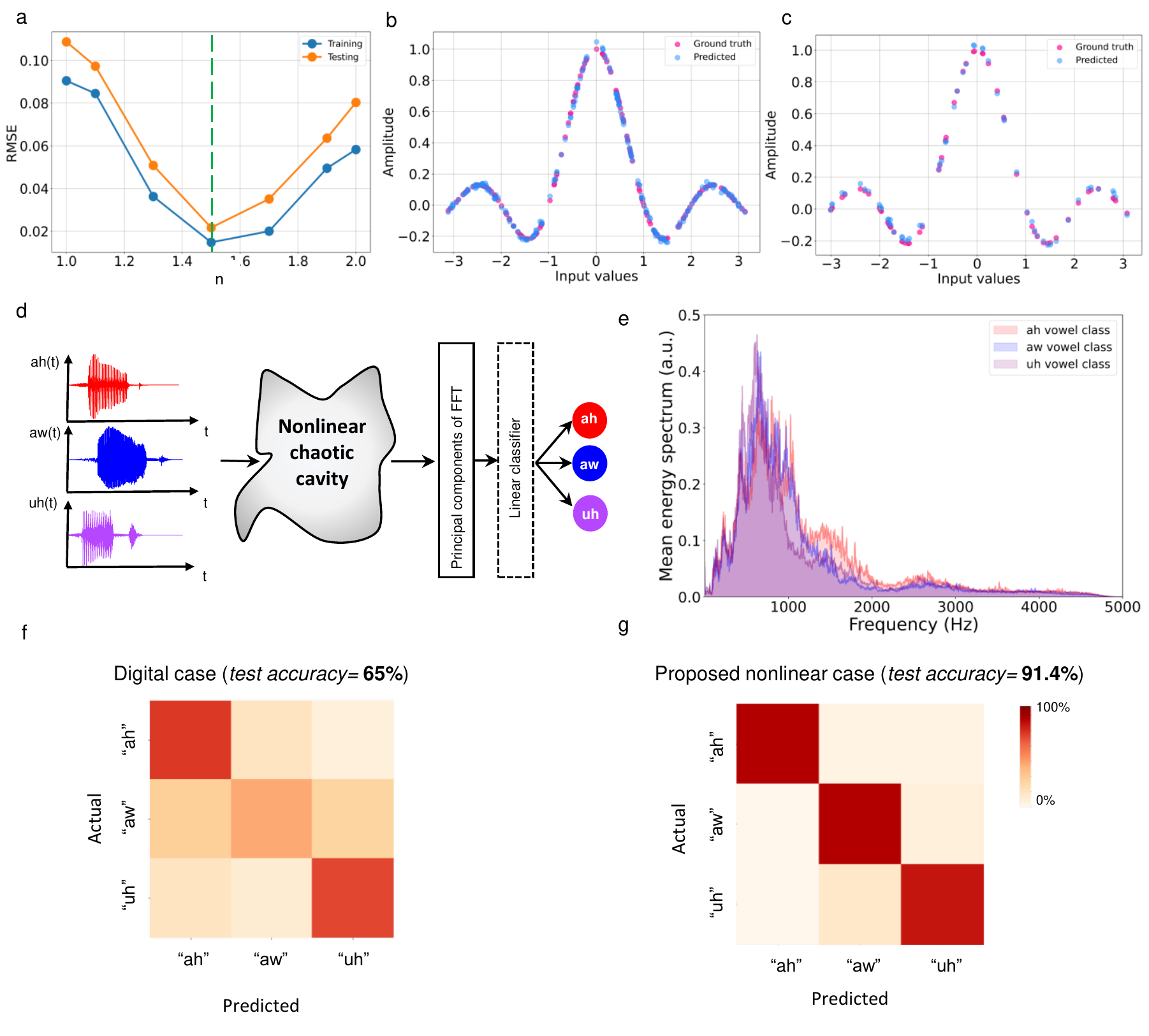}
	\caption{\textbf{ Performance after learning, for a nonlinear regression task and vowel recognition.} \textbf{a}   Tuning the nonlinear exponent $\textit{n}$ of nonlinear meta-scatterers allows us to reach a minimum RMSE for a standard regression learning task, namely interpolating the sinc function. \textbf{b} and \textbf{c} Training and testing regression performance for $\textit{n}=1.5$, respectively. \textbf{d} Schematic of the vowel recognition setup.  Raw audio waveforms of spoken vowel samples from three classes are independently injected via the source loudspeakers. After passing through a nonlinear chaotic cavity, the principal components of the FFT of the output signals are fed to a linear classifier.    \textbf{e} Frequency content of the "ah", "aw", and "uh" vowel classes. \textbf{f} and \textbf{g}  Confusion matrices and corresponding test accuracies for digital and nonlinear cases, respectively.
}\label{Fig_3}
\end{figure}

{T}he performance of the proposed physics-inspired computing system {can be illustrated} by starting with a simple regression problem, on a dataset generated with nonlinear relations. Linear regression of nonlinear functions is a standard and well-known analysis method to evaluate the learning ability of a neural network model which is impossible without a nonlinear transformation \cite{momeni2022electromagnetic}. 
The input information ($\zeta_{in}$) is a set of randomly generated numbers between $-\pi$ to $\pi$ and the corresponding output labels ($\textit{y}_i$) are generated according to nonlinear function $y=sinc(\zeta_{in})$.
Each input value ($\zeta_{in}$) is mapped to a matrix ($\mathbf{S}^{in}$) by multiplying it with a fixed random 2D mask, here of dimension M×N (N=M=10), where  $M$ and $N$ correspond to the numbers of loudspeakers and numbers of inputs.
We assume that each row of $\mathbf{S}^{in}$ is modulated at distinct close-by frequencies $\omega_n$, such that: 
\begin{equation}\label{e3}
\psi^{in}_{\textit{m}}(\textit{t})= \sum_{\textit{n}=1}^{N} \mathbf{S}^{in}_{\textit{n},\textit{m}} \sin(\omega_\textit{n}\textit{t}) 
  \end{equation}
The normalized $\psi^{in}_{\textit{m}}(\textit{t})$ is injected into cavity by the $\textit{m}${th} loudspeaker (see \textcolor{blue}{Fig. \ref{Fig_1}c}).
The nonlinear transformed waveforms are received by ten microphones. Linear regression is performed on the output, defined as the intensity of the second harmonic measured for each training input $\zeta_{in}$. The learning performance is measured by root-mean-squared error (RMSE). In order to achieve minimum RMSE for a regression learning task, we tune the nonlinear exponent $n$ of nonlinear meta-scatterers.  The regression task's RMSE values as functions of the hyperparameter $\textit{n}$ for training and testing dataset  are shown in \textcolor{blue}{Fig. \ref{Fig_3}a}. 
A remarkable learning performance, with a very low RMSE, is obtained for
$\textit{n}=1.5$. 
The regression results for the training and testing dataset by setting $\textit{n}$ to this optimal value are shown in \textcolor{blue}{Fig. \ref{Fig_3}b} and \textcolor{blue}{c}, respectively. These results demonstrate that the nonlinear data mapping occurring in the proposed computing system is sufficient to perform standard nonlinear regression tasks.
It should be noted that in order to compute the intensity of the second harmonics at the decision layer, we simply have to rescale the linear regression weights without any heavy additional computational cost (see Methods for further details).

\subsection{Vowel recognition }

{T}he proposed wave-based system can be trained directly on real sounds, without any pre-processing or encoding of the input. We thus turn to a more challenging task: the classification of spoken vowel sounds. To this end, the model is trained on a standard vowel recognition dataset \cite{hillenbrand1995acoustic,romera2018vowel}, while prediction accuracy is measured for new unseen data that are not used during training. We use a dataset consisting of 930 raw audio recordings of 10 vowel classes from 45 different male speakers and 48 different female speakers \cite{hillenbrand1995acoustic}. For our learning task, we select a subset of 279 recordings corresponding to three vowel classes that could not be classified by a linear classifier, represented by the vowel sounds "ah", "aw", and "uh" (see \textcolor{blue}{Fig. \ref{Fig_3}d}). To illustrate it, we plot the mean energy spectrum of these three vowel classes for all samples (male and female speakers)in \textcolor{blue}{Fig. \ref{Fig_3}e}.  There is a strong overlap between the spectral energy densities of these three vowel classes for the entire frequency range. Therefore, the chosen vowel recognition task is highly nontrivial.

The audio waveform of each vowel is injected in the ten loudspeakers on the left side of the cavity with 10ms time delay, between each loudspeaker, emitting waveforms that propagate through a disordered cavity (see Methods for further details). The waveforms are received by ten microphones on the location of nonlinear meta-scatterers. By recording the nonlinear transferred waveforms with microphones, principal components of the Fast Fourier Transform (FFT) of transformed signals are performed. Finally, we use a simple linear SVM to classify the vowel classes (see Methods for further details).

\begin{figure}[t!]
	\centering
	\includegraphics[width=\textwidth]{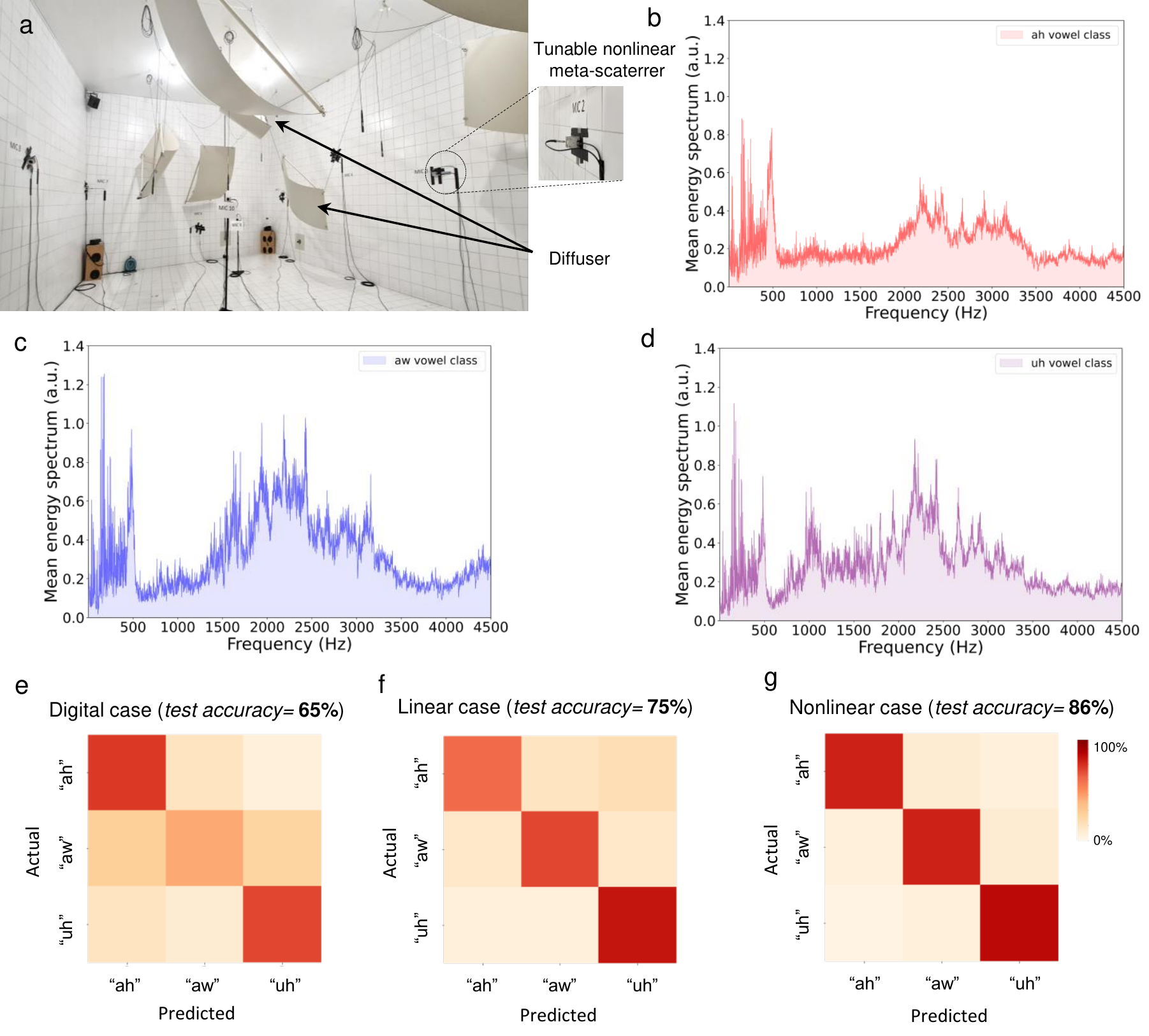}
	\caption{\textbf{Vowel recognition in a complex reverberation room.} \textbf{a}  Picture of the reverberation room and the installed setup. \textbf{b}-\textbf{d} The mean energy spectrum for each of the 
    three transformed vowel data (male and female speakers) normalized by the initial mean energy spectrum
    of data. \textbf{e}-\textbf{g} Testing accuracy of digital, linear (control off), and nonlinear (control on) cases, respectively.
}\label{Fig_4}
\end{figure}

The training results and corresponding confusion matrices for testing sets are shown in \textcolor{blue}{Fig. \ref{Fig_3}f-g}. {The quality of the vowel recognition for the test dataset is  compared using confusion matrices}. The confusion matrices (in this case, a 3×3 matrix) in this test scenario characterize the generalization ability of the network. We observe that the proposed sound-based system achieves excellent accuracy on unseen data. The observed test accuracy was $91.4\%$ which is comparable with state-of-the-art works such as physical recurrent neural networks \cite{hughes2019wave,qu2022resonance}.
The latter, however, requires highly expensive training on the paired input-output data, which is not the case of the proposed approach and has been only tested for three vowel classes that are already linearly separable in the frequency domain with a linear SVM (see supplementary information for further details). We  compared our results to state-of-the-art works as well as digital neural networks such as nonlinear SVM (see Supplementary Tables 1 and 2 in supplementary information).  
Some of these prior arts also used intrinsic material nonlinearity, which impairs energy efficiency.
On the other hand, the proposed wave-based computing system makes use of nonlinear meta-scatterers, which are power-efficient.
Furthermore, the proposed wave-based computing system is not only versatile and easy-to-train but also processes acoustic data in its native domain, working on real sounds without any pre-processing such as clipping or filtering.

In order to investigate the impact of the nonlinear mixing of sound waves in the cavity, we remove our nonlinear chaotic cavity from the block diagram in \textcolor{blue}{Fig. \ref{Fig_3}d} and calculate the test accuracy. In this case, the test accuracy drops to $65\%$ demonstrating the pivotal role of nonlinear wave mixing.

\subsection{Vowel recognition in realistic scenario}
Finally, we repeat the aforementioned experiment in a reverberation chamber to show the ability of the proposed method for vowel recognition in a real acoustic environment.
A reverberation chamber with additional acoustic diffusers is often used to create sound waves with statistically random incidence. Sound propagation in such an environment results in a multitude of multiple-scattering events that scramble the sound field, creating a speckle. We employ ten nonlinear meta-scatterers and microphones randomly located in the room (see \textcolor{blue}{Fig. \ref{Fig_4}a}). 
The audio waveform of each vowel is injected by a single loudspeaker located in the corner of the room. The transformed signals are recorded by microphones and the same training procedure as in  \textcolor{blue}{Fig. \ref{Fig_3}d} is carried out. \textcolor{blue}{Figs. \ref{Fig_4}b-d} show the  mean energy spectrum for each of the three transformed vowel data (male and female speakers) normalized by the initial mean energy spectrum of the data of \textcolor{blue}{Fig. \ref{Fig_3}e}. We notice that nonlinear multiple-scattering is very efficient at transforming these initially overlapping spectra into distinct sounds, particularly in two frequency ranges (50 to 600 Hz and 1000 Hz to 3500 Hz). As in the smaller cavity prototype, this enables significantly their classification by a linear classifier. 
The training results for different scenarios including digital case, linear case, and nonlinear case are shown in  \textcolor{blue}{Fig. \ref{Fig_3}e-g}, respectively. The confusion matrix result for the digital case is computed similarly to the previous section by removing the nonlinear chaotic cavity from the block diagram in \textcolor{blue}{Fig. \ref{Fig_3}d}, indicating the poor performance of a digital SVM. For a fair comparison, we report the confusion matrix results of linear and nonlinear cases, noting that the only difference between these two cases is that the nonlinear meta-scatterers are disabled in the linear case. The test accuracy is $75\%$ and $86\%$ for linear and nonlinear cases, respectively, confirming the importance of nonlinearities in the performance of the device.

In conclusion, We have shown that the dynamics of wave propagation in nonlinear multiple-scattering media are conceptually equivalent to reservoir computing systems. This conceptual connection opens up the opportunity for wave-based machine learning with efficient processing of information in its native domain. We have experimentally shown nonlinear multiple-scattering in a realistic acoustic system can be used to enable wave-based speech recognition, by allowing for strong and tailored nonlinear mapping of real sounds to a higher-dimensional space, without any pre-processing, and in a real-world environment.

\section{Methods}
\subsection{Experimental setup}
The cavity for performing reservoir computing in \textcolor{blue}{Fig. \ref{Fig_1}} to \textcolor{blue}{Fig. \ref{Fig_3}} is 2m in length, 1m in width, and 0.2m in height. It presents 11 modes below 500 Hz (see its memory characterization in Suppl. S3). For the vowel recognition task, the audio waveform of each vowel is injected by ten loudspeakers (with 10ms time delay between each loudspeaker) on the left side of the cavity. The experiments for the final vowel recognition in \textcolor{blue}{Fig. \ref{Fig_4}} are carried out in a real reverberant chamber. In this case, the waveform of each vowel is injected by one loudspeaker at the corner of the reverberant chamber. The nonlinear scatterers used in both configurations are randomly placed. They are implemented by controlling electrodynamic loudspeakers (Visaton FRWS 5 SC). The active controls are performed with an FPGA-based Speedgoat performance real-time target machine controlled by the xPC target environment of MATLAB (via SIMULINK). The output voltage is converted to a current through a Howland-Current Pump (HCP) converter \cite{etienne_IEEE}, with a conversion gain of 10mA/V, and is finally sent back to each controlled loudspeaker. The measurements (source generation and signal recording) are also carried out with the real-time target machine.

\subsection{Training of readout}
{Here, we show how to train the decision layer using the data of temporal signals received at the read-out nodes without using extra filtering operations. 
Consider $\psi^{out}_\textit{m}(\textit{t})$ as the temporal signal received at the output microphone $\textit{m}$, and its discrete Fourier transform of the discretized signal $\bar\psi^{out}_{\textit{m}}=\bigg(\psi^{out}_\textit{m}(0), \psi^{out}_\textit{m}(\textit{t}_0 ), … ,\psi^{out}_\textit{m}((\textit{n}-1)\textit{t}_0 )\bigg)^T$ defined by $\textit{y}_{f}=\sum_{(\textit{k}=0)}^{(\textit{n}-1)}\psi^{out}_\textit{m} (t_\textit{k} ) exp{(\frac{-2\pi i}{\textit{n}}\textit{j}\textit{k})}$ where $\textit{j}= 0, …, \textit{n}-1$, $(.)^T$ is the transpose operation, and $\textit{t}_0$ is the  sampling time. The relation between the Fourier coefficients  $y_{f_\textit{j}}$   and the discretized signal is described as the multiplication  $\mathbf{y}_{f}^\textit{m}=\mathbf{F}\bar\psi^{out}_\textit{m}$, where $\mathbf{F}$ is Fourier matrix:  
\begin{equation*}
\mathbf{F}= 
\begin{pmatrix}
1 & 1 & \cdots & 1 \\
1 & \kappa^2 & \cdots & \kappa^{\textit{n}-1} \\
\vdots  & \vdots  & \ddots & \vdots  \\
1 & \kappa^{\textit{n}-1} & \cdots & \kappa^{(\textit{n}-1)^2}
\end{pmatrix}
\end{equation*}
where $\kappa=e^{\frac{-2\pi i}{\textit{n}}}$. Generally, we are only interested in some frequency harmonics  whose component $\textit{m}$ can be calculated by  $\mathbf{y}_{f}^\textit{m}=\mathbf{F_s}\bar\psi^{out}_\textit{m}$, where $\mathbf{F_s}$ is the Fourier matrix corresponding to the desired frequency harmonics.
In order to train the readout function by linear regression (which is commonly defined by a linear matrix operation of the form $\mathbf{Y}={\mathbf{W}}\mathbf{X}^T$, where $\mathbf{W}$ is the weights matrix), we must compose both operations, multiplying the $\mathbf{F_s}$ and weight matrices:  

\begin{align}
\mathbf{Y}=\mathbf{W}\bigg({vec}\big(\mathbf{F_s}\mathbf{X}\big)\bigg)^T    
\end{align} 
where $\mathbf{X}=\bigg(\bar\psi^{out}_{1},\bar\psi^{out}_{2}, ... , \bar\psi^{out}_\textit{m}\bigg)$. vec($\mathbf{X}$) is the vectorization of a matrix $\mathbf{X}$  into a column vector by linear transformation and is described by $vec(\mathbf{X})=\sum_{\textit{i}=1}^{\textit{n}}\mathbf{e}_{\textit{i}} \bigotimes \mathbf{X}\mathbf{e}_{\textit{i}} $, where  $\mathbf{e}_{\textit{i}}$ is the $\textit{i}-th$ canonical basis vector for the n-dimensional space, that is $\mathbf{e}_{\textit{i}}= [ 0, ..., 0, 1, 0, ..., 0]^T$ and $\bigotimes$ is Kronecker product.
Similar to a regular linear regression ($\mathbf{Y}=\mathbf{W}\mathbf{X}^T$), the output of the proposed wave-based computing scheme involves a simple multiplication of matrices without sensitive or complex filters. This can also be viewed as a mere rescaling of the weight matrix of the digital layer. So, this comes with no extra computational cost.}

For learning the nonlinear function (sinc function), we used a supervised learning algorithm, linear regression, to train the readout function.  The predicted output is compared with the ground truth, and the error is calculated and used to update the weights in the readout network following the standard linear regression learning rule. 

 To train the readout network, for the vowel recognition  tasks, we used PyTorch, which provides a high-level application programming interface to access TensorFlow. According to Fig. 3d, firstly we performed PCA on normalized $\mathbf{y}_{f}^\textit{m}$ (we normalized (Min-Max normalization) the Fourier component of the data for two different ranges of 10-1kHz and 1-3.5kHz). PCA (Principal component analysis) is a linear dimensionality reduction method that uses Singular Value Decomposition of the data to project it to a lower dimensional space. Thereafter, we used linear SVM to classify vowel classes.

\textbf{Acknowledgements}:
A. Momeni and R. Fleury acknowledge funding from the Swiss National Science Foundation under the Eccellenza grant number 181232. X. Guo and H. Lissek acknowledge funding from the Swiss National Science Foundation (SNSF) under grant No. ${200020}\_{200498}$.

\textbf{Contributions}
A.M and R.F jointly conceived the idea. A.M and X.G. contributed equally. A.M. designed the computational engine and performed the theoretical, and numerical simulations. X.G. designed the nonlinear active controls and carried out the experiments. H.L. supervised the experimental work. R.F initiated and supervised the project. All authors contributed to the interpretation of the results and the writing of the manuscript.


\bibliography{cas-refs}

\begin{thebibliography}{10}
\expandafter\ifx\csname url\endcsname\relax
  \def\url#1{\texttt{#1}}\fi
\expandafter\ifx\csname urlprefix\endcsname\relax\def\urlprefix{URL }\fi
\expandafter\ifx\csname doiprefix\endcsname\relax\def\doiprefix{DOI }\fi
\providecommand{\bibinfo}[2]{#2}
\providecommand{\eprint}[2][]{\url{#2}}

\bibitem{gigan2022imaging}
\bibinfo{author}{Gigan, S.}
\newblock \bibinfo{journal}{\bibinfo{title}{Imaging and computing with
  disorder}}.
\newblock {\emph{\JournalTitle{Nature Physics}}} \textbf{\bibinfo{volume}{18}},
  \bibinfo{pages}{980--985} (\bibinfo{year}{2022}).

\bibitem{cao2022shaping}
\bibinfo{author}{Cao, H.}, \bibinfo{author}{Mosk, A.~P.} \&
  \bibinfo{author}{Rotter, S.}
\newblock \bibinfo{journal}{\bibinfo{title}{Shaping the propagation of light in
  complex media}}.
\newblock {\emph{\JournalTitle{Nature Physics}}} \textbf{\bibinfo{volume}{18}},
  \bibinfo{pages}{994--1007} (\bibinfo{year}{2022}).

\bibitem{zangeneh2021analogue}
\bibinfo{author}{Zangeneh-Nejad, F.}, \bibinfo{author}{Sounas, D.~L.},
  \bibinfo{author}{Al{\`u}, A.} \& \bibinfo{author}{Fleury, R.}
\newblock \bibinfo{journal}{\bibinfo{title}{Analogue computing with
  metamaterials}}.
\newblock {\emph{\JournalTitle{Nature Reviews Materials}}}
  \textbf{\bibinfo{volume}{6}}, \bibinfo{pages}{207--225}
  (\bibinfo{year}{2021}).

\bibitem{wright2022deep}
\bibinfo{author}{Wright, L.~G.} \emph{et~al.}
\newblock \bibinfo{journal}{\bibinfo{title}{Deep physical neural networks
  trained with backpropagation}}.
\newblock {\emph{\JournalTitle{Nature}}} \textbf{\bibinfo{volume}{601}},
  \bibinfo{pages}{549--555} (\bibinfo{year}{2022}).

\bibitem{feldmann2021parallel}
\bibinfo{author}{Feldmann, J.} \emph{et~al.}
\newblock \bibinfo{journal}{\bibinfo{title}{Parallel convolutional processing
  using an integrated photonic tensor core}}.
\newblock {\emph{\JournalTitle{Nature}}} \textbf{\bibinfo{volume}{589}},
  \bibinfo{pages}{52--58} (\bibinfo{year}{2021}).

\bibitem{silva2014performing}
\bibinfo{author}{Silva, A.} \emph{et~al.}
\newblock \bibinfo{journal}{\bibinfo{title}{Performing mathematical operations
  with metamaterials}}.
\newblock {\emph{\JournalTitle{Science}}} \textbf{\bibinfo{volume}{343}},
  \bibinfo{pages}{160--163} (\bibinfo{year}{2014}).

\bibitem{wang2022optical}
\bibinfo{author}{Wang, T.} \emph{et~al.}
\newblock \bibinfo{journal}{\bibinfo{title}{An optical neural network using
  less than 1 photon per multiplication}}.
\newblock {\emph{\JournalTitle{Nature Communications}}}
  \textbf{\bibinfo{volume}{13}}, \bibinfo{pages}{123} (\bibinfo{year}{2022}).

\bibitem{momeni2021reciprocal}
\bibinfo{author}{Momeni, A.} \emph{et~al.}
\newblock \bibinfo{journal}{\bibinfo{title}{Reciprocal metasurfaces for on-axis
  reflective optical computing}}.
\newblock {\emph{\JournalTitle{IEEE Transactions on Antennas and Propagation}}}
  \textbf{\bibinfo{volume}{69}}, \bibinfo{pages}{7709--7719}
  (\bibinfo{year}{2021}).

\bibitem{sol2022meta}
\bibinfo{author}{Sol, J.}, \bibinfo{author}{Smith, D.~R.} \&
  \bibinfo{author}{Del~Hougne, P.}
\newblock \bibinfo{journal}{\bibinfo{title}{Meta-programmable analog
  differentiator}}.
\newblock {\emph{\JournalTitle{Nature communications}}}
  \textbf{\bibinfo{volume}{13}}, \bibinfo{pages}{1--10} (\bibinfo{year}{2022}).

\bibitem{babaee2021parallel}
\bibinfo{author}{Babaee, A.}, \bibinfo{author}{Momeni, A.},
  \bibinfo{author}{Abdolali, A.} \& \bibinfo{author}{Fleury, R.}
\newblock \bibinfo{journal}{\bibinfo{title}{Parallel analog computing based on
  a 2$\times$ 2 multiple-input multiple-output metasurface processor with
  asymmetric response}}.
\newblock {\emph{\JournalTitle{Physical Review Applied}}}
  \textbf{\bibinfo{volume}{15}}, \bibinfo{pages}{044015}
  (\bibinfo{year}{2021}).

\bibitem{del2018leveraging}
\bibinfo{author}{del Hougne, P.} \& \bibinfo{author}{Lerosey, G.}
\newblock \bibinfo{journal}{\bibinfo{title}{Leveraging chaos for wave-based
  analog computation: demonstration with indoor wireless communication
  signals}}.
\newblock {\emph{\JournalTitle{Physical Review X}}}
  \textbf{\bibinfo{volume}{8}}, \bibinfo{pages}{041037} (\bibinfo{year}{2018}).

\bibitem{matthes2019optical}
\bibinfo{author}{Matth{\`e}s, M.~W.}, \bibinfo{author}{Del~Hougne, P.},
  \bibinfo{author}{De~Rosny, J.}, \bibinfo{author}{Lerosey, G.} \&
  \bibinfo{author}{Popoff, S.~M.}
\newblock \bibinfo{journal}{\bibinfo{title}{Optical complex media as universal
  reconfigurable linear operators}}.
\newblock {\emph{\JournalTitle{Optica}}} \textbf{\bibinfo{volume}{6}},
  \bibinfo{pages}{465--472} (\bibinfo{year}{2019}).

\bibitem{momeni2019generalized}
\bibinfo{author}{Momeni, A.}, \bibinfo{author}{Rajabalipanah, H.},
  \bibinfo{author}{Abdolali, A.} \& \bibinfo{author}{Achouri, K.}
\newblock \bibinfo{journal}{\bibinfo{title}{Generalized optical signal
  processing based on multioperator metasurfaces synthesized by susceptibility
  tensors}}.
\newblock {\emph{\JournalTitle{Physical Review Applied}}}
  \textbf{\bibinfo{volume}{11}}, \bibinfo{pages}{064042}
  (\bibinfo{year}{2019}).

\bibitem{momeni2022switchable}
\bibinfo{author}{Momeni, A.}, \bibinfo{author}{Rouhi, K.} \&
  \bibinfo{author}{Fleury, R.}
\newblock \bibinfo{journal}{\bibinfo{title}{Switchable and simultaneous
  spatiotemporal analog computing with computational graphene-based
  multilayers}}.
\newblock {\emph{\JournalTitle{Carbon}}} \textbf{\bibinfo{volume}{186}},
  \bibinfo{pages}{599--611} (\bibinfo{year}{2022}).

\bibitem{anderson2023optical}
\bibinfo{author}{Anderson, M.~G.}, \bibinfo{author}{Ma, S.-Y.},
  \bibinfo{author}{Wang, T.}, \bibinfo{author}{Wright, L.~G.} \&
  \bibinfo{author}{McMahon, P.~L.}
\newblock \bibinfo{journal}{\bibinfo{title}{Optical transformers}}.
\newblock {\emph{\JournalTitle{arXiv preprint arXiv:2302.10360}}}
  (\bibinfo{year}{2023}).

\bibitem{wetzstein2020inference}
\bibinfo{author}{Wetzstein, G.} \emph{et~al.}
\newblock \bibinfo{journal}{\bibinfo{title}{Inference in artificial
  intelligence with deep optics and photonics}}.
\newblock {\emph{\JournalTitle{Nature}}} \textbf{\bibinfo{volume}{588}},
  \bibinfo{pages}{39--47} (\bibinfo{year}{2020}).

\bibitem{wu2020neuromorphic}
\bibinfo{author}{Wu, Z.}, \bibinfo{author}{Zhou, M.}, \bibinfo{author}{Khoram,
  E.}, \bibinfo{author}{Liu, B.} \& \bibinfo{author}{Yu, Z.}
\newblock \bibinfo{journal}{\bibinfo{title}{Neuromorphic metasurface}}.
\newblock {\emph{\JournalTitle{Photonics Research}}}
  \textbf{\bibinfo{volume}{8}}, \bibinfo{pages}{46--50} (\bibinfo{year}{2020}).

\bibitem{lin2018all}
\bibinfo{author}{Lin, X.} \emph{et~al.}
\newblock \bibinfo{journal}{\bibinfo{title}{All-optical machine learning using
  diffractive deep neural networks}}.
\newblock {\emph{\JournalTitle{Science}}} \textbf{\bibinfo{volume}{361}},
  \bibinfo{pages}{1004--1008} (\bibinfo{year}{2018}).

\bibitem{sebastian2020memory}
\bibinfo{author}{Sebastian, A.}, \bibinfo{author}{Le~Gallo, M.},
  \bibinfo{author}{Khaddam-Aljameh, R.} \& \bibinfo{author}{Eleftheriou, E.}
\newblock \bibinfo{journal}{\bibinfo{title}{Memory devices and applications for
  in-memory computing}}.
\newblock {\emph{\JournalTitle{Nature nanotechnology}}}
  \textbf{\bibinfo{volume}{15}}, \bibinfo{pages}{529--544}
  (\bibinfo{year}{2020}).

\bibitem{markovic2020physics}
\bibinfo{author}{Markovi{\'c}, D.}, \bibinfo{author}{Mizrahi, A.},
  \bibinfo{author}{Querlioz, D.} \& \bibinfo{author}{Grollier, J.}
\newblock \bibinfo{journal}{\bibinfo{title}{Physics for neuromorphic
  computing}}.
\newblock {\emph{\JournalTitle{Nature Reviews Physics}}}
  \textbf{\bibinfo{volume}{2}}, \bibinfo{pages}{499--510}
  (\bibinfo{year}{2020}).

\bibitem{grollier2020neuromorphic}
\bibinfo{author}{Grollier, J.} \emph{et~al.}
\newblock \bibinfo{journal}{\bibinfo{title}{Neuromorphic spintronics}}.
\newblock {\emph{\JournalTitle{Nature electronics}}}
  \textbf{\bibinfo{volume}{3}}, \bibinfo{pages}{360--370}
  (\bibinfo{year}{2020}).

\bibitem{pierangeli2019large}
\bibinfo{author}{Pierangeli, D.}, \bibinfo{author}{Marcucci, G.} \&
  \bibinfo{author}{Conti, C.}
\newblock \bibinfo{journal}{\bibinfo{title}{Large-scale photonic ising machine
  by spatial light modulation}}.
\newblock {\emph{\JournalTitle{Physical review letters}}}
  \textbf{\bibinfo{volume}{122}}, \bibinfo{pages}{213902}
  (\bibinfo{year}{2019}).

\bibitem{huang2021antiferromagnetic}
\bibinfo{author}{Huang, J.}, \bibinfo{author}{Fang, Y.} \&
  \bibinfo{author}{Ruan, Z.}
\newblock \bibinfo{journal}{\bibinfo{title}{Antiferromagnetic spatial photonic
  ising machine through optoelectronic correlation computing}}.
\newblock {\emph{\JournalTitle{Communications Physics}}}
  \textbf{\bibinfo{volume}{4}}, \bibinfo{pages}{242} (\bibinfo{year}{2021}).

\bibitem{suykens1999least}
\bibinfo{author}{Suykens, J.~A.} \& \bibinfo{author}{Vandewalle, J.}
\newblock \bibinfo{journal}{\bibinfo{title}{Least squares support vector
  machine classifiers}}.
\newblock {\emph{\JournalTitle{Neural processing letters}}}
  \textbf{\bibinfo{volume}{9}}, \bibinfo{pages}{293--300}
  (\bibinfo{year}{1999}).

\bibitem{seber2003linear}
\bibinfo{author}{Seber, G.~A.} \& \bibinfo{author}{Lee, A.~J.}
\newblock \emph{\bibinfo{title}{Linear regression analysis}}, vol.
  \bibinfo{volume}{330} (\bibinfo{publisher}{John Wiley \& Sons},
  \bibinfo{year}{2003}).

\bibitem{rafayelyan2020large}
\bibinfo{author}{Rafayelyan, M.}, \bibinfo{author}{Dong, J.},
  \bibinfo{author}{Tan, Y.}, \bibinfo{author}{Krzakala, F.} \&
  \bibinfo{author}{Gigan, S.}
\newblock \bibinfo{journal}{\bibinfo{title}{Large-scale optical reservoir
  computing for spatiotemporal chaotic systems prediction}}.
\newblock {\emph{\JournalTitle{Physical Review X}}}
  \textbf{\bibinfo{volume}{10}}, \bibinfo{pages}{041037}
  (\bibinfo{year}{2020}).

\bibitem{pierangeli2021photonic}
\bibinfo{author}{Pierangeli, D.}, \bibinfo{author}{Marcucci, G.} \&
  \bibinfo{author}{Conti, C.}
\newblock \bibinfo{journal}{\bibinfo{title}{Photonic extreme learning machine
  by free-space optical propagation}}.
\newblock {\emph{\JournalTitle{Photonics Research}}}
  \textbf{\bibinfo{volume}{9}}, \bibinfo{pages}{1446--1454}
  (\bibinfo{year}{2021}).

\bibitem{teugin2021scalable}
\bibinfo{author}{Te{\u{g}}in, U.}, \bibinfo{author}{Y{\i}ld{\i}r{\i}m, M.},
  \bibinfo{author}{O{\u{g}}uz, {\.I}.}, \bibinfo{author}{Moser, C.} \&
  \bibinfo{author}{Psaltis, D.}
\newblock \bibinfo{journal}{\bibinfo{title}{Scalable optical learning
  operator}}.
\newblock {\emph{\JournalTitle{Nature Computational Science}}}
  \textbf{\bibinfo{volume}{1}}, \bibinfo{pages}{542--549}
  (\bibinfo{year}{2021}).

\bibitem{pathak2018model}
\bibinfo{author}{Pathak, J.}, \bibinfo{author}{Hunt, B.},
  \bibinfo{author}{Girvan, M.}, \bibinfo{author}{Lu, Z.} \&
  \bibinfo{author}{Ott, E.}
\newblock \bibinfo{journal}{\bibinfo{title}{Model-free prediction of large
  spatiotemporally chaotic systems from data: A reservoir computing approach}}.
\newblock {\emph{\JournalTitle{Physical review letters}}}
  \textbf{\bibinfo{volume}{120}}, \bibinfo{pages}{024102}
  (\bibinfo{year}{2018}).

\bibitem{momeni2022electromagnetic}
\bibinfo{author}{Momeni, A.} \& \bibinfo{author}{Fleury, R.}
\newblock \bibinfo{journal}{\bibinfo{title}{Electromagnetic wave-based extreme
  deep learning with nonlinear time-floquet entanglement}}.
\newblock {\emph{\JournalTitle{Nature communications}}}
  \textbf{\bibinfo{volume}{13}}, \bibinfo{pages}{1--11} (\bibinfo{year}{2022}).

\bibitem{yildirim2022nonlinear}
\bibinfo{author}{Yildirim, M.} \emph{et~al.}
\newblock \bibinfo{journal}{\bibinfo{title}{Nonlinear optical data transformer
  for machine learning}}.
\newblock {\emph{\JournalTitle{arXiv preprint arXiv:2208.09398}}}
  (\bibinfo{year}{2022}).

\bibitem{oguz2022programming}
\bibinfo{author}{Oguz, I.} \emph{et~al.}
\newblock \bibinfo{journal}{\bibinfo{title}{Programming nonlinear propagation
  for efficient optical learning machines}}.
\newblock {\emph{\JournalTitle{arXiv preprint arXiv:2208.04951}}}
  (\bibinfo{year}{2022}).

\bibitem{vandoorne2014experimental}
\bibinfo{author}{Vandoorne, K.} \emph{et~al.}
\newblock \bibinfo{journal}{\bibinfo{title}{Experimental demonstration of
  reservoir computing on a silicon photonics chip}}.
\newblock {\emph{\JournalTitle{Nature communications}}}
  \textbf{\bibinfo{volume}{5}}, \bibinfo{pages}{3541} (\bibinfo{year}{2014}).

\bibitem{zhou2021large}
\bibinfo{author}{Zhou, T.} \emph{et~al.}
\newblock \bibinfo{journal}{\bibinfo{title}{Large-scale neuromorphic
  optoelectronic computing with a reconfigurable diffractive processing unit}}.
\newblock {\emph{\JournalTitle{Nature Photonics}}}
  \textbf{\bibinfo{volume}{15}}, \bibinfo{pages}{367--373}
  (\bibinfo{year}{2021}).

\bibitem{hughes2019wave}
\bibinfo{author}{Hughes, T.~W.}, \bibinfo{author}{Williamson, I.~A.},
  \bibinfo{author}{Minkov, M.} \& \bibinfo{author}{Fan, S.}
\newblock \bibinfo{journal}{\bibinfo{title}{Wave physics as an analog recurrent
  neural network}}.
\newblock {\emph{\JournalTitle{Science advances}}}
  \textbf{\bibinfo{volume}{5}}, \bibinfo{pages}{eaay6946}
  (\bibinfo{year}{2019}).

\bibitem{skinner1994optical}
\bibinfo{author}{Skinner, S.~R.}, \bibinfo{author}{Steck, J.~E.} \&
  \bibinfo{author}{Behrman, E.~C.}
\newblock \bibinfo{title}{Optical neural network using kerr-type nonlinear
  materials}.
\newblock In \emph{\bibinfo{booktitle}{Proceedings of the fourth international
  conference on microelectronics for neural networks and fuzzy systems}},
  \bibinfo{pages}{12--15} (\bibinfo{organization}{IEEE}, \bibinfo{year}{1994}).

\bibitem{rumelhart1986learning}
\bibinfo{author}{Rumelhart, D.~E.}, \bibinfo{author}{Hinton, G.~E.} \&
  \bibinfo{author}{Williams, R.~J.}
\newblock \bibinfo{journal}{\bibinfo{title}{Learning representations by
  back-propagating errors}}.
\newblock {\emph{\JournalTitle{nature}}} \textbf{\bibinfo{volume}{323}},
  \bibinfo{pages}{533--536} (\bibinfo{year}{1986}).

\bibitem{zhong2021dynamic}
\bibinfo{author}{Zhong, Y.} \emph{et~al.}
\newblock \bibinfo{journal}{\bibinfo{title}{Dynamic memristor-based reservoir
  computing for high-efficiency temporal signal processing}}.
\newblock {\emph{\JournalTitle{Nature communications}}}
  \textbf{\bibinfo{volume}{12}}, \bibinfo{pages}{408} (\bibinfo{year}{2021}).

\bibitem{sun2021sensor}
\bibinfo{author}{Sun, L.} \emph{et~al.}
\newblock \bibinfo{journal}{\bibinfo{title}{In-sensor reservoir computing for
  language learning via two-dimensional memristors}}.
\newblock {\emph{\JournalTitle{Science advances}}}
  \textbf{\bibinfo{volume}{7}}, \bibinfo{pages}{eabg1455}
  (\bibinfo{year}{2021}).

\bibitem{marinella2019efficient}
\bibinfo{author}{Marinella, M.~J.} \& \bibinfo{author}{Agarwal, S.}
\newblock \bibinfo{journal}{\bibinfo{title}{Efficient reservoir computing with
  memristors}}.
\newblock {\emph{\JournalTitle{Nature Electronics}}}
  \textbf{\bibinfo{volume}{2}}, \bibinfo{pages}{437--438}
  (\bibinfo{year}{2019}).

\bibitem{liu2022optoelectronic}
\bibinfo{author}{Liu, K.} \emph{et~al.}
\newblock \bibinfo{journal}{\bibinfo{title}{An optoelectronic synapse based on
  $\alpha$-in2se3 with controllable temporal dynamics for multimode and
  multiscale reservoir computing}}.
\newblock {\emph{\JournalTitle{Nature Electronics}}}
  \textbf{\bibinfo{volume}{5}}, \bibinfo{pages}{761--773}
  (\bibinfo{year}{2022}).

\bibitem{du2017reservoir}
\bibinfo{author}{Du, C.} \emph{et~al.}
\newblock \bibinfo{journal}{\bibinfo{title}{Reservoir computing using dynamic
  memristors for temporal information processing}}.
\newblock {\emph{\JournalTitle{Nature communications}}}
  \textbf{\bibinfo{volume}{8}}, \bibinfo{pages}{2204} (\bibinfo{year}{2017}).

\bibitem{milano2022materia}
\bibinfo{author}{Milano, G.} \emph{et~al.}
\newblock \bibinfo{journal}{\bibinfo{title}{In materia reservoir computing with
  a fully memristive architecture based on self-organizing nanowire networks}}.
\newblock {\emph{\JournalTitle{Nature Materials}}}
  \textbf{\bibinfo{volume}{21}}, \bibinfo{pages}{195--202}
  (\bibinfo{year}{2022}).

\bibitem{moon2019temporal}
\bibinfo{author}{Moon, J.} \emph{et~al.}
\newblock \bibinfo{journal}{\bibinfo{title}{Temporal data classification and
  forecasting using a memristor-based reservoir computing system}}.
\newblock {\emph{\JournalTitle{Nature Electronics}}}
  \textbf{\bibinfo{volume}{2}}, \bibinfo{pages}{480--487}
  (\bibinfo{year}{2019}).

\bibitem{tanaka2019recent}
\bibinfo{author}{Tanaka, G.} \emph{et~al.}
\newblock \bibinfo{journal}{\bibinfo{title}{Recent advances in physical
  reservoir computing: A review}}.
\newblock {\emph{\JournalTitle{Neural Networks}}}
  \textbf{\bibinfo{volume}{115}}, \bibinfo{pages}{100--123}
  (\bibinfo{year}{2019}).

\bibitem{etienne_IEEE}
\bibinfo{author}{{Rivet}, E.}, \bibinfo{author}{{Karkar}, S.} \&
  \bibinfo{author}{{Lissek}, H.}
\newblock \bibinfo{journal}{\bibinfo{title}{Broadband low-frequency
  electroacoustic absorbers through hybrid sensor-/shunt-based impedance
  control}}.
\newblock {\emph{\JournalTitle{IEEE Trans. Control Syst. Technol.}}}
  \textbf{\bibinfo{volume}{25}}, \bibinfo{pages}{63--72}
  (\bibinfo{year}{2017}).
\newblock \doiprefix 10.1109/TCST.2016.2547981.

\bibitem{GUO_JSV_2022}
\bibinfo{author}{Guo, X.}, \bibinfo{author}{Volery, M.} \&
  \bibinfo{author}{Lissek, H.}
\newblock \bibinfo{journal}{\bibinfo{title}{Pid-like active impedance control
  for electroacoustic resonators to design tunable single-degree-of-freedom
  sound absorbers}}.
\newblock {\emph{\JournalTitle{J. Sound Vib.}}} \textbf{\bibinfo{volume}{525}},
  \bibinfo{pages}{116784} (\bibinfo{year}{2022}).
\newblock \doiprefix https://doi.org/10.1016/j.jsv.2022.116784.

\bibitem{GUO_PRApplied_2020}
\bibinfo{author}{Guo, X.}, \bibinfo{author}{Lissek, H.} \&
  \bibinfo{author}{Fleury, R.}
\newblock \bibinfo{journal}{\bibinfo{title}{Improving sound absorption through
  nonlinear active electroacoustic resonators}}.
\newblock {\emph{\JournalTitle{Phys. Rev. Applied}}}
  \textbf{\bibinfo{volume}{13}}, \bibinfo{pages}{014018}
  (\bibinfo{year}{2020}).
\newblock \doiprefix 10.1103/PhysRevApplied.13.014018.

\bibitem{hillenbrand1995acoustic}
\bibinfo{author}{Hillenbrand, J.}, \bibinfo{author}{Getty, L.~A.},
  \bibinfo{author}{Clark, M.~J.} \& \bibinfo{author}{Wheeler, K.}
\newblock \bibinfo{journal}{\bibinfo{title}{Acoustic characteristics of
  american english vowels}}.
\newblock {\emph{\JournalTitle{The Journal of the Acoustical society of
  America}}} \textbf{\bibinfo{volume}{97}}, \bibinfo{pages}{3099--3111}
  (\bibinfo{year}{1995}).

\bibitem{romera2018vowel}
\bibinfo{author}{Romera, M.} \emph{et~al.}
\newblock \bibinfo{journal}{\bibinfo{title}{Vowel recognition with four coupled
  spin-torque nano-oscillators}}.
\newblock {\emph{\JournalTitle{Nature}}} \textbf{\bibinfo{volume}{563}},
  \bibinfo{pages}{230--234} (\bibinfo{year}{2018}).

\bibitem{qu2022resonance}
\bibinfo{author}{Qu, Y.}, \bibinfo{author}{Zhou, M.}, \bibinfo{author}{Khoram,
  E.}, \bibinfo{author}{Yu, N.} \& \bibinfo{author}{Yu, Z.}
\newblock \bibinfo{journal}{\bibinfo{title}{Resonance for analog recurrent
  neural network}}.
\newblock {\emph{\JournalTitle{ACS Photonics}}} \textbf{\bibinfo{volume}{9}},
  \bibinfo{pages}{1647--1654} (\bibinfo{year}{2022}).

\end{thebibliography}
\newpage

\textbf {\LARGE Supplementary Information for:}
\newline
\newline
\centering \textbf{\large   Physics-inspired Neuroacoustic Computing Based on Tunable Nonlinear Multiple-scattering}
\newline
\newline
\author{Ali Momeni,}
\author{Xinxin Guo,}
\author{Hervé Lissek, and }
\author{Romain Fleury}
\affil[1]{Laboratory of Wave Engineering, Ecole Polytechnique Fédérale de Lausanne (EPFL), CH-1015 Lausanne, Switzerland.}
\affil[2]{Signal Processing Laboratory LTS2, Ecole Polytechnique Fédérale de Lausanne (EPFL), CH-1015 Lausanne, Switzerland}
\affil[*]{E-mail: romain.fleury@epfl.ch}

\raggedright
\textbf{This PDF file includes:}

Section S1. Derivation of the sound wave equation as a reservoir computing update equations 

Section S2. Nonlinear active scatterers

Section S3. Memory: impulse responses of the employed cavity and reverberation chamber

Section S4. Comparison of the proposed computing system with state-of-the-art works and digital neural networks

References 
\newpage
\raggedright
{\textbf{Section S1. Derivation of the sound wave equation as a reservoir computing update equations} }

As we mentioned in the main text, the dynamics of a scalar wave field distribution, $\textit{p}( \textit{x}, \textit{y}, \textit{z})$, are governed by the wave equation:
\begin{equation}\tag{S1}
\frac{\partial^{2}\textit{p}}{\partial \textit{t}^{2}}-\textit{c}^{2} \cdot \nabla^{2} \textit{p}=\psi^{in}_\textit{t}
  \end{equation}
We can discretize the (S1) by centered finite differences with a temporal step size of $\Delta_\textit{t}$: 
\begin{equation}\tag{S2}
\frac{\textit{p}_{\textit{t}+1}-2p_\textit{t}+\textit{p}_{t-1}}{\Delta_t^2}-c^{2} \cdot \nabla^{2} \textit{p}_t=\psi^{in}_\textit{t}
  \end{equation}
The subscript $t$ is used to indicate the value of a scalar field at a given time step. We rewrite (S2) in matrix form as
\begin{equation}\label{e2-annex}
\mathbf{h}_\textit{t}=\mathbf{Q}^{(c)}(\mathbf{h}_{t-1}) \cdot \mathbf{h}_{t-1}+\mathbf{Q}^{(i)} \cdot \mathbf{x}_\textit{t}
  \end{equation}
where $\mathbf{h}_\textit{t}=\begin{pmatrix}
 \textit{p}_{\textit{t}+1} \\
\textit{p}_\textit{t} 
\end{pmatrix}$. Also $\mathbf{Q}^{(c)}$ and $\mathbf{Q}^{(i)}$ are defined as
\begin{align}\tag{S4}\nonumber
&\mathbf{Q}^{(c)}=\begin{pmatrix}
 2+ \Delta_t^2 \cdot \textit{c}^2 \cdot \nabla^{2}& -1\\
1 & 0
\end{pmatrix}\\ \nonumber
&\mathbf{Q}^{(i)}=\Delta_t^2 \begin{pmatrix}
 \mathbf{G}^{(i)} \\
\mathbf{0}
\end{pmatrix}\\ \nonumber
  \end{align}
 where $\mathbf{0}$ is a matrix of all zeros. The output of the system can be expressed as 
\begin{align}\tag{S5} 
 \psi^{out}=\bigg|[{\mathbf{G}^{(o)}}^T ,0] \cdot \mathbf{h}_\textit{t} \bigg|
 \end{align}
where $\mathbf{G}^{(i)}$ and $\mathbf{G}^{(o)}$ are the respective spatial distributions of the injection and measurement points in the wave propagation domain.

\raggedright
{\textbf{Section S2. Nonlinear active scatterers } }

The feedback control loop shown in \textcolor{blue}{Supplementary Fig. \ref{Fig_5}b} is considered to implement the desired nonlinear scatterers. The pressure in front of each loudspeaker ($p_f$) is sensed in real-time and used as the control input. The control output (in voltage) is defined in the specified form of $(\textit{t})=\textit{G}_{NL}\textit{p}_f(\textit{t})^\textit{n}$ and is converted to the feedback current. The index $\textit{n}$ is set to $1.5$ which is optimal for the two vowel recognition tasks considered. The constant control gain $\textit{G}_{NL}$ is tuned to trigger the nonlinear effect under low input power. The excitation source is imposed to remain $\textit{p}_f$ in the range of a few pascals so that the loudspeaker behaves linearly in the passive case (control off). Therefore, nonlinearity can only be provided when active control is applied to the loudspeakers. 

\begin{figure}[htbp]
	\centering \setcounter{figure}{0}
	\includegraphics[width=\textwidth]{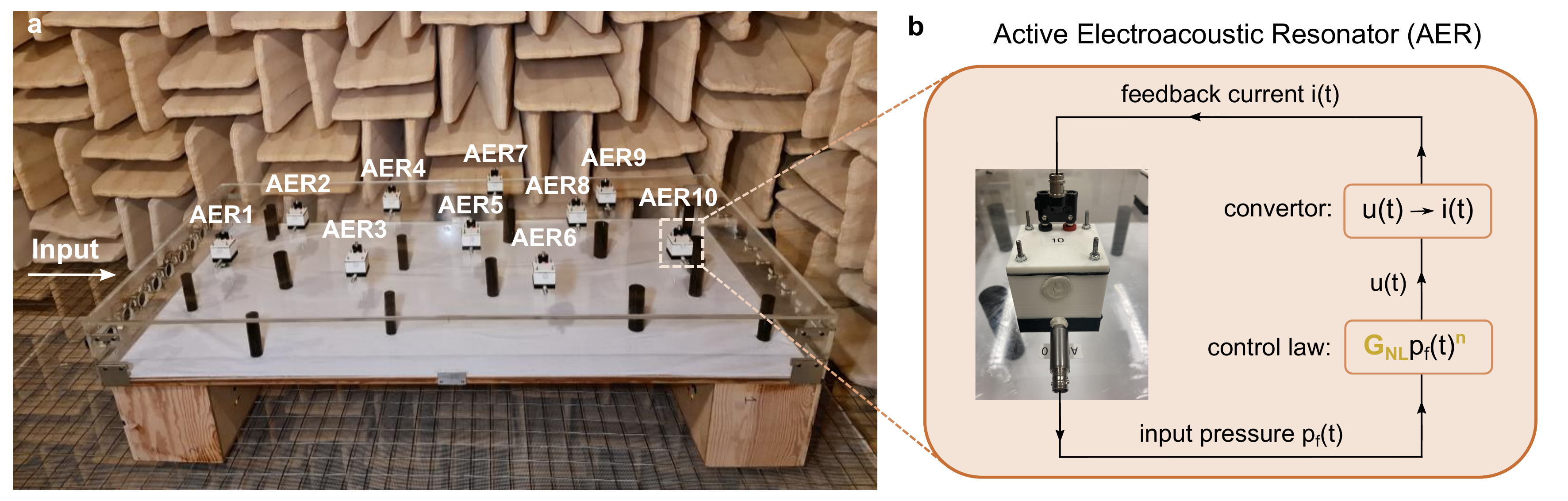}\renewcommand{\figurename}{Supplementary Figure}
	\caption{\textbf{Arrangement of the nonlinear active scatterers on the top side of the cavity.} \textbf{a} Locations of the ten scatterers being controlled. \textbf{b} Nonlinear active control applied.
}\label{Fig_5}
\end{figure}

\begin{figure}[htbp]
	\centering 
	\includegraphics[width=\textwidth]{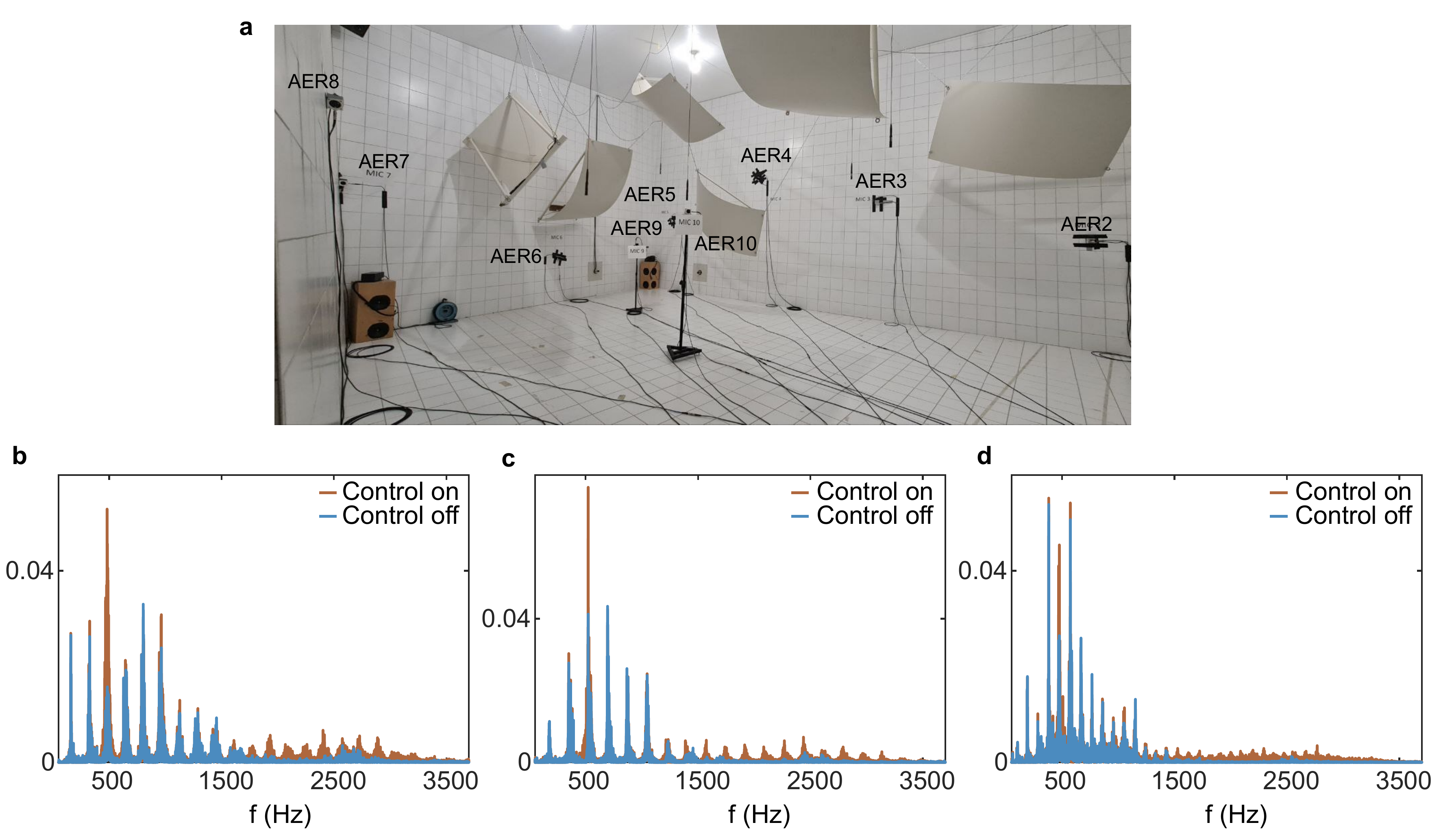}\renewcommand{\figurename}{Supplementary Figure}
	\caption{\textbf{Arrangement of the nonlinear active scatterers in the reverberant room.} \textbf{a} Locations of the scatterers being controlled in the reverberant room. \textbf{b} \textbf{c} \textbf{d} Comparison between control off (linear) and control on (nonlinear) cases. The illustrated results are the transmitted pressures measured by the microphone near AER1 when one of the vowels of each class (\textbf{b}: class ah, \textbf{c}: class aw, \textbf{d}: class uh) is considered as the source.
}\label{Fig_6}
\end{figure}

The settings of the control gain $G_{NL}$ are summarized in Table 1, for the two vowel recognition tasks, respectively. The corresponding numbering of the loudspeakers is given in \textcolor{blue}{Supplementary Fig. \ref{Fig_5}a} and \textcolor{blue}{Supplementary Fig. \ref{Fig_6}a-b}, respectively.

{\centering
\fontsize{5}{3}
\small
\begin{tabular}{ |p{4cm}||p{4cm}|p{8cm} | }
 \hline
 \multicolumn{3}{|c|}{Supplementary Table 1: Parameter settings of nonlinear active controls } \\
 \hline
\textbf{ Tasks }& \textbf{Scatterers } &\textbf{Setting details}\\
 \hline
 \textbf{V recognition in the cavity }   & 
AER1 to AER10 (index shown in \textcolor{blue}{Fig. \ref{Fig_5}a})

& $G_{NL}= 7.0$ for AER1

$G_{NL}= 5.0$ for AER2

$G_{NL}= 6.0$ for AER3

$G_{NL}= 6.8$ for AER4

$G_{NL}= 7.0$ for AER5

$G_{NL}= 5.0$ for AER6

$G_{NL}= 6.0$ for AER7

$G_{NL}= 6.3$ for AER8

$G_{NL}= 6.0$ for AER9

$G_{NL}= 5.8$ for AER10
\\

  \hline
 
\textbf{Vowel recognition in the real reverberant room} &
AER1 to AER10 (index shown in \textcolor{blue}{Fig. \ref{Fig_6}a})

 & $G_{NL}= 9.0$ for AER1

$G_{NL}= 6.4$ for AER2

$G_{NL}= 5.4$ for AER3

$G_{NL}= 7.9$ for AER4

$G_{NL}= 7.5$ for AER5

$G_{NL}= 6.4$ for AER6

$G_{NL}= 6.9$ for AER7

$G_{NL}= 7.1$ for AER8

$G_{NL}= 7.1$ for AER9

$G_{NL}= 9.8$ for AER10 
\\

 \hline
    \end{tabular}}

\newpage
For a better illustration of the triggered nonlinear effects, we show in \textcolor{blue}{Supplementary Fig. \ref{Fig_6}c-e} the examples of transmitted vowel signals for each class. A noticeable difference can be identified between the linear (control off) and the nonlinear (control on) cases. The considered three classes of vowels become distinguishable only when the nonlinearity is introduced, as evidenced in \textcolor{blue}{Fig. \ref{Fig_4}b-d} in the main manuscript.

\raggedright
{\textbf{Section S3. Memory: impulse responses of the employed cavity and reverberate chamber} }

The memory effect can be directly visualized by the impulse response, i.e., the output of the proposed computing system when excited by the Dirac delta
function $\psi^{in}(\textit{t}) = \delta(\textit{t}-\textit{t}_0)$. The pulse is located at time step $t_0$. The memory effect is illustrated by the fact that the impact of input at time step $t_0$ extends into a certain time horizon. Here, the proposed cavity and room are excited by a 0.1 ms pulse signal, in order to measure the temporal responses at different readout nodes (see Supplementary Figures 1 and 2). As clear from Supplementary Figures 1 and 2, the amplitude and time evolution of temporal impulse responses are different for each readout node, showing distinct temporal mapping with different lifetimes.

\begin{figure}
	\centering
	\includegraphics[width=\textwidth]{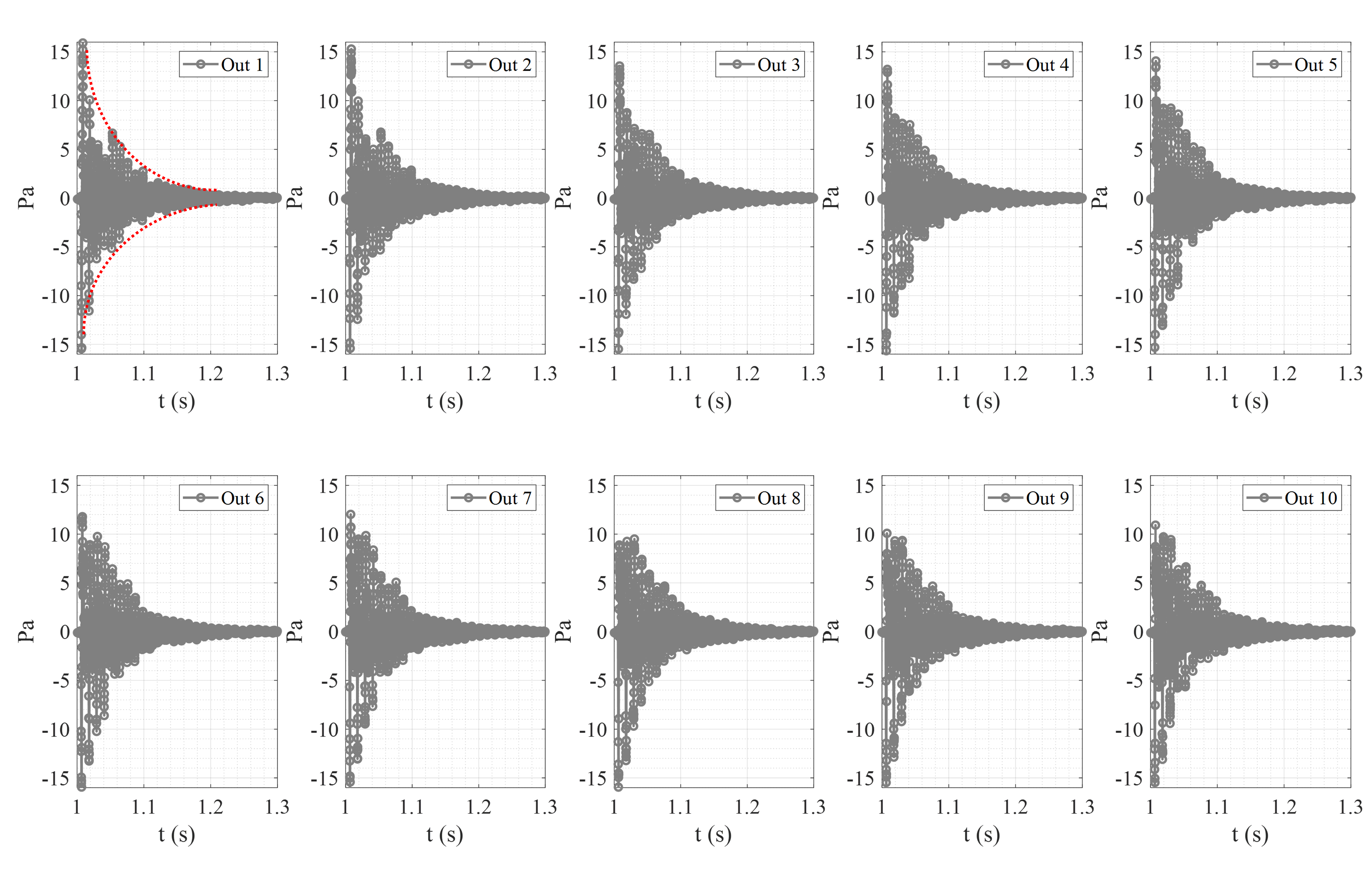}\renewcommand{\figurename}{Supplementary Figure}
	\caption{\textbf{Impulse responses of the nonlinear chaotic cavity}. The impulse responses of the nonlinear chaotic cavity at different readout nodes when it is excited by a 1 ms pulse at $t=1s$.
}\label{Fig_4}
\end{figure}
\begin{figure}
	\centering
	\includegraphics[width=\textwidth]{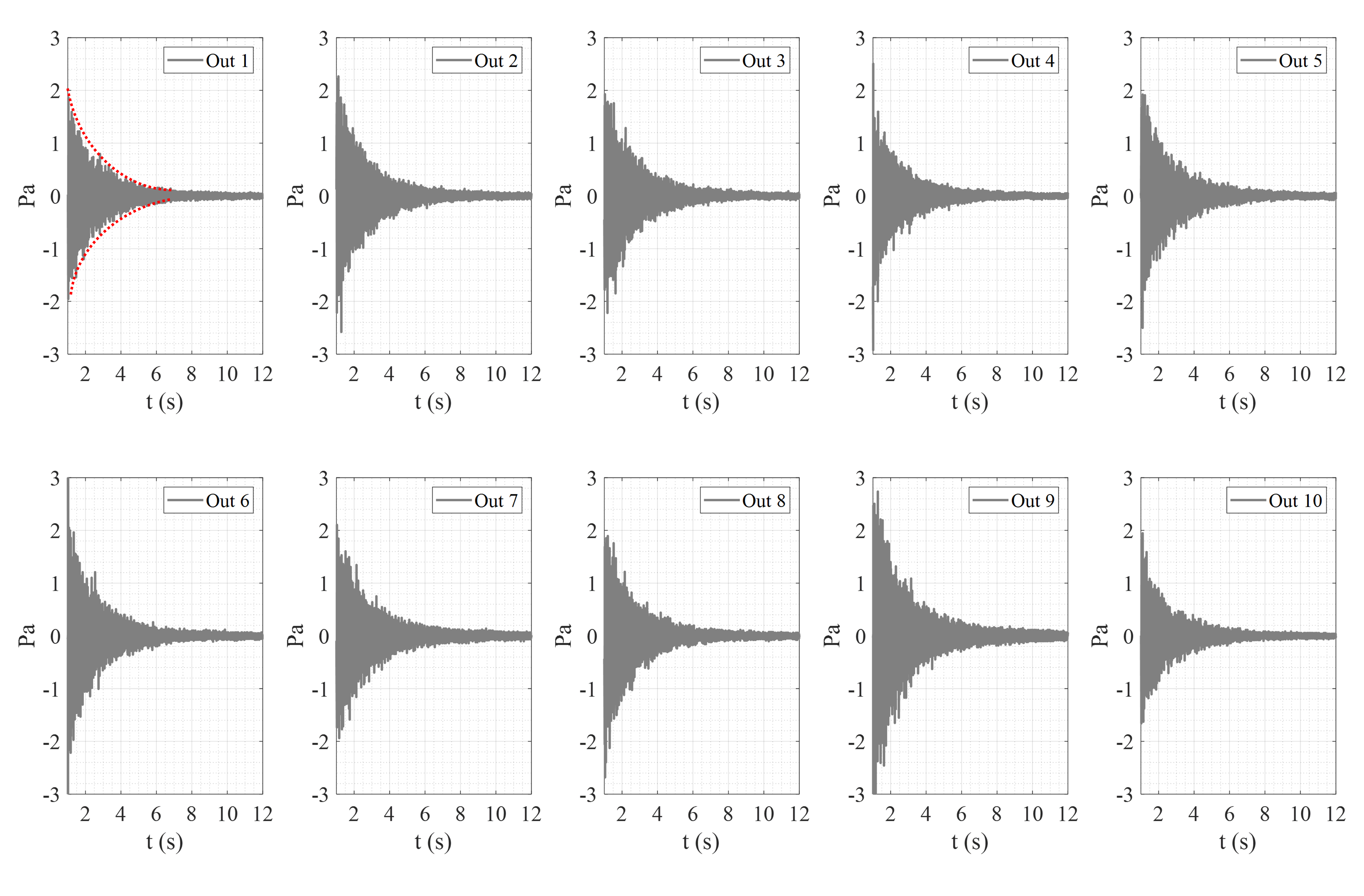} \renewcommand{\figurename}{Supplementary Figure}
	\caption{\textbf{Impulse responses of the reverberation room.}. The impulse responses of the reverberation room at different readout nodes when it is excited by a 1 ms pulse at $t=1s$.
}\label{Fig_4}
\end{figure}

\raggedright
{\textbf{Section S4. Comparison of the proposed computing system with state-of-the-art works and digital neural networks} }

\textcolor{black}{Here, we compare the performance of the proposed computing system to prior works as well as digital neural networks. The accuracy and RMSE comparison between the computation architectures are presented in Supplementary Tables 2 and 3.}

{\centering
\fontsize{5}{3}
\small
\begin{tabular}{ |p{4cm}||p{4cm}|p{8cm} | }
 \hline
 \multicolumn{3}{|c|}{Supplementary Table 2: Comparison of the proposed computing system with prior works } \\
 \hline
\textbf{ Tasks}& \textbf{Benchmark} &\textbf{Details}\\
 \hline
 \textbf{Learning nonlinear function ($\textbf{sinc(\textit{x})}$)}   & 
Ref [\cite{teugin2021scalable}]: RMSE: 0.0671

Our work: RMSE: 0.017

& • Ref [\cite{teugin2021scalable}] used a GRIN 50/125 multimode fiber (MMF) that supports 240 spatial modes. In this work, the authors used a high-power optical pulse in order to excite the nonlinearity of GRIN MMF (input optical peak power pulse equal to 3.43kW). \\

  \hline
 
\textbf{Vowel recognition} &
Ref [\cite{qu2022resonance}]: Acc.: 83.3$\%$ 

Ref [\cite{hughes2019wave}]: Acc.: 86.3$\%$ ~~~~~

Our work: Acc.: 91.4$\%$

 &
• Refs [\cite{qu2022resonance}] and [\cite{hughes2019wave}] employed wave-based recurrent neural networks. They used ae, ei, and iy vowel classes that are linearly separable in the frequency domain (90.3\% test accuracy with a Linear SVM and 200 trainable parameters in the frequency domain). Ref[\cite{hughes2019wave}] used 4200 trainable parameters.  

• We used ah, aw, and uh vowel classes (65.0\% test accuracy with a Linear SVM and only 200 trainable parameters).  The corresponding vowel recognition for such classes is nontrivial.
\\

 \hline
    \end{tabular}}

{\centering
\fontsize{5}{3}
\small
\begin{tabular}{ |p{4.5cm}||p{5cm}|p{7cm} | }
 \hline
 \multicolumn{3}{|c|}{Supplementary Table 3: Comparison between digital neural networks and the proposed wave-based system } \\
 \hline
\textbf{ Network Structure}& \textbf{Accuracy on vowel dataset (\%)} &\textbf{Details}\\
 \hline
 \textbf{ Linear SVM }  & 
Test Accuracy= 65 ± 0.5

& Linear SVM with 200 trainable parameters. \\

  \hline
 
\textbf{Nonlinear  SVM (sigmoid)} &
Test Accuracy= 68± 1    ~~~~~

 &
Nonlinear  SVM with the sigmoid nonlinear kernel and 200 trainable parameters.
\\
 
 \hline
 \textbf{Nonlinear  SVM (RBF)} &

Test Accuracy= 74± 1    ~~~~~

 &
Nonlinear SVM with the Radial Basis Function nonlinear kernel and 200 trainable parameters.

\\
 
 \hline
  \textbf{ Recurrent neural network} &

Test Accuracy= 80.2    ~~~~~

 &
Digital Recurrent neural network with Leaky ReLU nonlinearity and 5250 trainable parameters \cite{hughes2019wave}.

\\
 
 \hline
\textbf{Proposed  system} &

Test Accuracy= 91.4    ~~~~~

 &
Only 200 trainable parameters with a linear SVM
at decision layer\\
 
 \hline
    \end{tabular}}

\end{document}